\documentclass[fleqn,twoside]{article}

\usepackage{espcrc2,psfig,epsfig}

\usepackage[figuresright]{rotating}

\input macros.sty

\newcommand{\AmS}{{\protect\the\textfont2
  A\kern-.1667em\lower.5ex\hbox{M}\kern-.125emS}}

\title{
\vspace{-4.4cm}
       \begin{flushright}
       {\normalsize
        \tt BI-TP 2004/26 \\
            CPT-2004/P.046\\
            DESY 04-171   \\
            IFIC/04-48    \\
            MPP-2004-108  \\}
       \end{flushright}
       \vspace{1.2cm}
  Correlation functions at small quark masses with overlap
  fermions\thanks{Talk presented by H. Wittig at
        {\it Lattice 2004}, Fermilab, 21--26 June 2004.}}

\author{L. Giusti\address{Centre de Physique Th\'eorique,
        CNRS Luminy, F-13288 Marseille Cedex 9, France},
        P. Hern\'andez\address{Dpto. F\'isica Te\'orica and IFIC,
        Edificio Institutos Investigaci\'on, E-46071 Valencia, Spain},
        M. Laine\address{Faculty of Physics, University of Bielefeld,
        D-33501 Bielefeld, Germany},
	C. Pena\address[DESY]{Deutsches Elektronen-Synchrotron, DESY,
        Notkestr. 85, D-22603 Hamburg, Germany},
	P. Weisz\address{Max-Planck-Institut f\"ur Physik, F\"ohringer
        Ring 6, D-80805 Munich, Germany},
	J. Wennekers\addressmark[DESY]
	and
	H. Wittig\addressmark[DESY]}
       
\begin{document}

\begin{abstract}
We report on recent work on the determination of low-energy constants
describing $\Delta{S}=1$ weak transitions, in order to investigate the
origins of the $\Delta{I}=1/2$ rule. We focus on numerical techniques
designed to enhance the statistical signal in three-point correlation
functions computed with overlap fermions near the chiral limit.
\end{abstract}

\maketitle

\section{INTRODUCTION}

Much recent activity in lattice QCD has focused on the determination
of low-energy constants (LECs), which parameterise non-perturbative
physics in an effective description at low energies. A promising
approach involves computing in the $\epsilon$--regime, the kinematical
region of arbitrarily small quark masses $m$ in a finite volume $V$
such that $m\Sigma V\;\lesssim\;1$. Here discretisations that preserve
chiral symmetry at non-zero lattice spacing should be used so that
small masses and momenta can be reached safely. However, simulations
in the $\epsilon$-regime are expensive: owing to spontaneous chiral
symmetry breaking the spectrum of the Dirac operator becomes
arbitrarily dense near the origin, and thus the operator is
ill-conditioned. Furthermore, one observes large statistical
fluctuations (``spikes'') in correlation functions of local currents
\cite{Bietenholz03,lma} for small masses. A number of efficient
numerical tools have been developed \cite{numeps,DeG_Sch04,lma}, in
which the low modes of the Dirac operator are treated exactly. In
ref.\,\cite{lma} we proposed a method, dubbed ``low-mode averaging''
(LMA), which was shown to eliminate the spikes in two-point
correlation functions for quark masses
$m\;\lesssim\;1/{\Sigma}V$. Here we report on the extension of LMA to
three-point correlation functions, in order to study non-leptonic kaon
decays.

\section{LMA REVISITED}

We consider the massive Neuberger operator \cite{NeubergerDirac}
\be
   D_m = (1-\half{\abar}m)D+m,\quad \abar=a/(1+s),
\ee
where $|s|<1$ is a free parameter and the massless Neuberger operator,
$D$, satisfies the Ginsparg-Wilson relation. We are interested in
correlation functions of the left-handed flavour current
\be
   [J_\mu]_{\alpha\beta}=(\psibar_\alpha\gamma_\mu P_{-}
   \psitilde_\beta), \quad P_{\pm}=\half(1\pm\gamma_5),
\ee
where $\psitilde\equiv(1-{\abar}D/2)\psi$, and $\alpha, \beta$ are
flavour indices. Using the left-handed current has the advantage that
zero modes of the Dirac operator are projected out in correlation
functions. However, modes corresponding to small non-zero eigenvalues
can still cause large statistical fluctuations. Let $S(x,y)$ denote
the quark propagator from $y$ to $x$. LMA works by separating $n_{\rm
low}$ low-lying modes and using the spectral representation of the
propagator. In this way the quark propagator is written as
\be
S(x,y)= \sum_{k=1}^{n_{\rm low}} \frac{e_k(x)\otimes
  e_k(y)^\dagger}{\alpha_k} +S^h(x,y),
\label{eq_Sxy}
\ee
where $S^h$ is the propagator in the orthogonal complement of the
subspace spanned by the $n_{\rm low}$ lowest modes. The vector $e_k$
is given by
\be
   e_k = P_\sigma u_k+ P_{-\sigma}D P_\sigma u_k,
\ee
where $-\sigma$ denotes the chirality of the sector containing zero
modes (if any), and $u_k$ is an approximate eigenmode of $D_m^\dagger
D_m$:
\be
   P_\sigma D_m^\dagger D_m P_\sigma u_k = \alpha_k u_k +r_k,\quad
   (u_k,r_i)=0.
\ee
After inserting the rhs. of \eq{eq_Sxy} into the expression for the
correlator $C(t)\equiv\sum_{\xvec}\langle J_0(x) J_0(0)\rangle$, one
picks up three contributions:
\be
   C(t) = C^{ll}(t)+C^{hl}(t)+C^{hh}(t).
\ee
As explained in \cite{lma}, the statistical signal for the correlator
is enhanced by exploiting translational invariance in $C^{ll}(t)$ and
$C^{hl}(t)$, so that these contributions are sampled over many
different source points. Applications of LMA to the calculation of
$F_\pi$ in the $\epsilon$-regime can be found in \cite{lma}.

\section{$K\to\pi\pi$ WITH ACTIVE CHARM}

In a recent paper \cite{strat} we proposed a strategy to investigate
the origins of the $\Delta{I}=1/2$ rule in $K\to\pi\pi$ decays,
i.e. the observed enhancement of the transition amplitude for a
two-pion final state with isospin~0, $|A_0|/|A_2|=22.1$.

The goal of our programme is to quantify separately the contributions
to the $\Delta{I}=1/2$ rule from physics at the charm quark mass scale
and from ``intrinsic'' QCD effects due to soft-gluon exchange. A
crucial part of our strategy is to keep an active charm quark, so that
the theory has an approximate $\fourby$ chiral symmetry. The
$\Delta{S}=1$ effective weak Hamiltonian after the Operator Product
Expansion contains the four-quark operators ${\cal Q}_1^{+}$ and
${\cal Q}_1^{-}$ which are given by
\bea
 & &\hspace{-0.65cm}{\cal Q}_1^{\pm}= (\sbar\gamma_\mu P_{-}\tilde{u})
                   (\ubar\gamma_\mu P_{-}\tilde{d}) {\pm}
                   (\sbar\gamma_\mu P_{-}\tilde{d})
                   (\ubar\gamma_\mu P_{-}\tilde{u}) \nonumber \\
 & &\hspace{0.5cm} - (u\to c). \label{eq_Q1pm_def}
\eea
${\cal Q}_1^+$ and ${\cal Q}_1^-$ are singlets under $\rm SU(4)_R$ and
transform under irreducible representations of $\rm SU(4)_L$ of
dimensions 84 and 20, respectively.

In Chiral Perturbation Theory (ChPT), the ratio of amplitudes
$|A_0|/|A_2|$ is, at leading order, related to a ratio of LECs via
\be
  \frac{|A_0|}{|A_2|} = 
  \frac{1}{\sqrt{2}}\left(\frac{1}{2}+\frac{3}{2}\frac{g_1^{-}}{g_1^{+}}
  \right).
\ee
Here, the LECs $g_1^\pm$ multiply the counterparts of the four-quark
operators ${\cal{Q}}_1^\pm$ in the low-energy theory. They can be
determined by studying suitable correlation functions of
${\cal{Q}}_1^\pm$ in QCD and matching them to the corresponding
expressions in ChPT. The fact that in the $\epsilon$-regime this
matching can be performed at NLO without the presence of additional
interaction terms with unknown coefficients \cite{strat}, makes this
an attractive setting for carrying out our programme.

The intrinsic QCD contributions to the enhancement can be isolated by
determining $g_1^{-}/g_1^{+}$ in the theory with
$m_u=m_d=m_s=m_c$. Then, by studying the ratio of LECs as $m_c$
departs from the mass-degenerate limit, the specific contribution of
the charm quark to the $\Delta{I}=1/2$ rule can be investigated in
detail.

The complicated renormalisation and mixing patterns of four-fermion
operators usually encountered in lattice formulations can be avoided
through the use of Ginsparg-Wilson fermions.
Indeed, the use of the modified fields $\psitilde=(1-{\abar}D/2)\psi$
in \eq{eq_Q1pm_def} guarantees that the renormalisation and mixing of
${\cal{Q}}_1^\pm$ are like in the continuum theory \cite{strat}. In
particular, no mixings with lower-dimensional operators can occur.
For full details we refer the reader to ref.~\cite{strat}. An
investigation of the effects of the charm quark based on ChPT has been
published in \cite{HerLai04}.


We now report on our attempts to extract the ratio $g_1^{-}/g_1^{+}$
in the SU(4)-symmetric theory. In this case only ``figure-8''-graphs
must be considered, since ``eye''-graphs cancel exactly for
$m_c=m_u$. We define the following three-point functions of
${\cal{Q}}_1^\pm$ and the left-handed flavour current:
\be
  C_1^\pm(x_0,y_0) = \sum_{\xvec,\yvec} \left\langle [J_0(x)]_{du}
  [{\cal{Q}}_1^\pm(0)] [J_0(y)]_{us} \right\rangle.
\ee
The relation between $g_1^{-}/g_1^{+}$ and the ratio of correlators is
given by
\be
  \frac{g_1^{-}}{g_1^{+}}\,H(x_0,y_0) =
  \frac{k_1^{-}}{k_1^{+}}\,\left. 
  \frac{C_1^{-}(x_0,y_0)}{C_1^{+}(x_0,y_0)} \right|_{\rm ren},
\label{eq_g1pm_R2}
\ee
where the factor $H(x_0,y_0)$ has been computed in ChPT in the
$\epsilon$-regime at NLO \cite{strat}. The factors $k_1^\pm$ are
Wilson coefficients, and it is assumed that $C_1^{-}/C_1^{+}$ is
renormalised in some scheme. From \eq{eq_g1pm_R2} one sees that the
ratio of correlators is directly proportional to the ratio of LECs.

\begin{figure}[t]
\begin{center}
\psfig{file=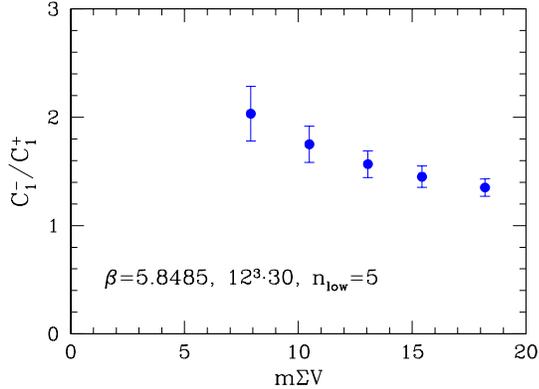,width=7.cm}
\vspace{-0.9cm}
\caption{The fitted value of the ratio $C_1^{-}/C_1^{+}$ plotted
  versus $m{\Sigma}V$ for $n_{\rm low}=5$. We have set
  $\Sigma=(250\,\MeV)^3$. \label{fig_20over84}} 
\end{center}
\vspace{-1.2cm}
\end{figure}

Our main results were obtained from 638 configurations on a lattice of
size $12^3\cdot30$ at $\beta=5.8485$, using the 5 lowest modes in the
implementation of LMA for figure-8 graphs. In Fig.\,\ref{fig_20over84}
we plot the ratio $C_1^{-}/C_1^{+}$ as a function of the quark mass in
units of $\Sigma V$. It is seen that $C_1^{-}/C_1^{+}$ increases as
the quark mass is tuned towards smaller values, which, at first sight,
might be interpreted as an enhancement in the ratio of LECs from QCD
effects alone. However, the quark masses used in the simulations are
too heavy to lie inside the $\epsilon$-regime, so that the computed
NLO matching factor $H(x_0,y_0)$ cannot be applied to yield
$g_1^{-}/g_1^{+}$. For smaller masses the statistical signal for the
ratio of three-point functions was lost even with LMA.

\begin{figure}[t]
\begin{center}
\psfig{file=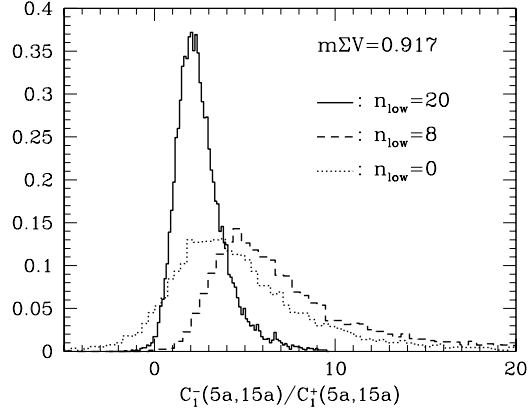,width=7.cm}
\vspace{-0.9cm}
\caption{Bootstrap distributions obtained for $m{\Sigma}V=0.917$
  at $\beta=5.8$ on a $8^3\cdot20$ lattice. \label{fig_boot}}
\end{center}
\vspace{-0.95cm}
\end{figure}

In order to demonstrate the feasibility of our strategy one has to
find ways of obtaining a strong signal in the $\epsilon$-regime. We
have thus investigated the quality of the signal for three-point
functions using LMA with a larger number of low modes
\cite{Kpipi_inprep}. Preliminary results on a lattice of size
$8^3\cdot20$ at $\beta=5.8$ are shown in Fig. \ref{fig_boot}, where we
have plotted the bootstrap distributions for the ratio of three-point
functions $C_1^{-}/C_1^{+}$. The plot shows that using $n_{\rm low}=8$
hardly improves the signal at all, while for $n_{\rm low}=20$ the
distribution is much narrower, so that the statistical signal is
indeed enhanced. We conclude that LMA leads to an improvement also for
three-point functions, but only if $n_{\rm low}$ is increased relative
to the previously studied case of two-point correlators. Obviously,
using larger values of $n_{\rm low}$ leads to a computational
overhead, which, however, is outweighed by far through the observed
enhancement of the signal. We are currently running on larger spatial
volumes in order to corroborate these findings.
\smallskip
\par\noindent{\bf Acknowledgements:}
Simulations were performed on PC clusters at DESY Hamburg, LRZ Munich
and University of Valencia. L.G. was supported by the EU under contract
HPRN-CT-2002-00311 (EURIDICE). P.H. was supported by the CICYT
(FPA2002-00162) and by the Generalitat Valenciana (CTIDIA/2002/5).


\end{document}